\documentstyle[twocolumn,twoside,prl,floats,aps]{revtex}
\tighten

\newcommand{\eq}{\begin{equation}}
\newcommand{\en}{\end{equation}}
\newcommand{\eqa}{\begin{eqnarray}}
\newcommand{\ena}{\end{eqnarray}}
\newcommand{\eqm}{\begin{mathletters}}
\newcommand{\enm}{\end{mathletters}}

\begin{document}
\draft

\vskip 0.4cm

\title{ A new universality class describing the insulating regime of
disordered wires with strong spin-orbit scattering}

\author{M. Caselle}

\address{  Dipartimento di Fisica
Teorica dell'Universit\`a di Torino and
 I.N.F.N., Sezione di Torino\\
 via P.Giuria 1, I-10125 Turin,Italy\\
\parbox{14 cm}{\medskip\rm\indent
We discuss the distribution of transmission eigenvalues in the strongly
localized regime in the presence of both a magnetic field and
spin--orbit scattering. We show that, under suitable conditions,
 this distribution can be described
by a new universality class labelled not only by the index $\beta$
but also by a new index $\eta$.
This result is obtained  by mapping the problem into that of a suitable
Calogero-Sutherland model. \\
\\
PACS numbers: 72.10.Bg, 05.60.+w, 72.15.Rn, 72.20.My }}
\maketitle
\narrowtext

Magnetoconductance (MC) in the strongly localized regime of mesoscopic
systems has been the subject of an increasing interest during last
years~\cite{mk}~-\cite{amww}.
While the same phenomenon in the weakly localized regime is by
now rather well understood, and a good agreement has been reached among
different theoretical descriptions and between their predictions and the
experimental results, the situation is less clear in the strongly
localized regime. In particular it is not clear to which extent the
Random Matrix Theory (RMT) of quantum transport~\cite{im,mps}, and the
corresponding classification in terms of universality classes,
originally developed to describe the weakly localized regime,
can  also be extended to the insulating regime. Indeed, it seems that
two different insulating regimes exist. The first one (strongly
insulating regime) in which the
localization length $\xi$ (to be defined below) is much smaller than the
hopping length $t_h$ ,  seems to be well described by a
directed-path approach (see ref.\cite{mk,amww} and references therein).
The second one (weakly insulating regime), close to the
Anderson transition, in which $t_h/\xi\sim 1$ and backscattering paths
are important, seems to be well
described within a RMT framework. This would be an important result
due to the simplicity and generality of the RMT
approach: the input required is only the symmetry of the Hamiltonian and
the resulting classification is completely encoded by a single index
$\beta$. The above scenario is  however complicated by the presence of
some conflicting theoretical analyses. The disagreement between the
various approaches becomes evident when looking at the MC behaviour
in presence of strong spin orbit scattering (SOS).
In fact in the insulating regime  the localization length is related
to the conductance by: $g\equiv \exp(-2L/\xi)$, ($L$=length of the
sample). In the standard RMT approach $\xi$ is  proportional to
$\beta$ and, as a consequence,
the magnetoconductance is directly related to the
the change of $\beta$ as a function of the magnetic field.
The RMT approach based on the maximal entropy ansatz
predicts that
the effect of a magnetic field on a sample with SOS is to change $\beta$
from $\beta=4$ (``symplectic ensamble'': spin-rotation symmetry broken
but time reversal symmetry conserved) to $\beta=2$ (``unitary
ensamble'': time reversal
symmetry broken, status of the spin-rotation symmetry irrelevant).
Defining as $\xi(H)$ the localization length in
presence of a magnetic field, from the above analysis it follows that
$\xi(H)/\xi=1/2$, and we expect
a strong {\sl negative} MC~\cite{pssd}.
On the contrary an exact calculation within the non-linear $\sigma$
model~\cite{el}
shows that if the spin-rotation symmetry is broken, $\beta$ remains
equal to
4 even in presence of a breaking of the time reversal symmetry
(thus implying a
zero  MC) and that in this case the ensamble behaves as unitary
($\beta=2$) only with respect to the level statistics.
In this letter we shall try to clarify this issue. Let us state our main
results:

{a]}\hskip0.5cm
 The main point we want to stress is that, while in the weakly localized
regime all the relevant physical properties are completely determined
by the level statistics  (namely the value of $\beta$),
 when extending the RMT approach to the insulating regime, due to the
fact that in this case {\it the conductance is  dominated by the lowest
eigenvalue} all the details of the chosen RMT model
(number of degrees of freedom, possibly the presence of new critical
indices) become important.

{b]}\hskip 0.5cm
In particular this gives a way to reconciliate the above described
controversy. We shall show below
that also within the maximal entropy approach a zero MC should be
expected in the insulating regime.

{c]}\hskip 0.5cm
It is however very interesting to notice that in
 some recent experiments~\cite{sohs} on samples characterized by
strong SOS  a clear negative MC has been observed in the weakly
insulating regime, in contradiction with the above results.
We shall  show  below that  under
suitable hypothesis, it does exist a new universality class in the RMT
description of quantum transport which indeed admits negative MC in the
insulating regime.
This new class has $\beta=4$, as
in~\cite{el}, but a new
index $\eta$ exists that allows to distinguish this case from the
ordinary symplectic ensamble. As a consequence of this new $\eta$ index
a negative MC is expected even if $\beta=4$.
\vskip 0.2cm

To show these results let us first briefly remind few known features of
the RMT description of quantum transport.
The central object in this approach is the $2N\times 2N$
 Transfer Matrix $M$ (where $N$ is the number of
scattering channels at the Fermi level)
which relates the incoming and outgoing fluxes in
a two-probe geometry (namely a finite disordered section of length
$L$ and transverse width $L_t$, to which current is supplied by two
semi-infinite ordered leads):
\eq
M\left(\begin{array}{c}I \\  O \end{array}\right)=
\left(\begin{array}{c}O' \\  I' \end{array}\right)~~~,
\en
where the I,O,I',O' vectors describe the incoming and outgoing wave
amplitudes on the left and right respectively.
To construct the RMT one must first impose the physical
requirements which define the ensamble of the matrices: for instance
flux conservation, or time
conjugation, if time reversal symmetry is conserved. This defines the
group $G$ to which the random matrix belongs. Then one assumes that the
physics of the problem only depends on a subset of the degrees of
freedom of the matrix.
It turns out that a good parametrization in our case is given
by the $N$ real, positive eigenvalues
$\lambda_i$ of the
matrix $Q=\frac{1}{4}[M^\dagger M +(M^\dagger M)^{-1} -2]$. These
parameters have an important physical interpretation since they are
related to the transmission eigenvalues $T_i$ by
$\lambda_{i}\equiv (1-T_{i})/T_{i}$ in terms of which
all the interesting physical
properties
 can be evaluated. For instance the conductance
$g$ is given by $g=g_0\sum_n T_n$
(with $g_0=2e^2/h$). This parametrization allows to identify a
subgroup $H$ of invariance in our
ensamble. The matrix ensamble  is thus related to a suitable $G/H$
coset. It turns out that in all the relevant cases $G/H$ is actually a
symmetric space.  The last step is the
identification of a dynamical principle to describe the ``time''
evolution of the eigenvalues. This will allow to construct explicitly
the probability distribution
 $P(\{\lambda_i\})$ for the eigenvalues $\lambda_i$.
 In the case of RMT for quantum transport
the ``time'' $s=L/l$ is actually the length of the conductor $L$
measured in unities of the mean free path $l$; the dynamical principle
is the so called ``local maximum entropy principle'' which simply
implies the free motion of the eigenvalues in the symmetric space $G/H$
and, as a consequence, the evolution operator is the radial part of the
Laplace-Beltrami operator on $G/H$. The corresponding (coupled)
differential equation for the probability distribution $P(\lambda)$ of
the eigenvalues is known as the  Dorokhov-Mello-Pereyra-Kumar (DMPK)
 equation~\cite{dmpk}
\begin{equation}
\frac{\partial P}{\partial s}~=
\frac{2}{\gamma}\sum_{i=1}^{N}
\frac{\partial}{\partial\lambda_{i}}\lambda_{i}(1+\lambda_{i})
J\frac{\partial}{\partial\lambda_{i}}J^{-1}~P
\equiv~D~P,
\label{DMPK}
\end{equation}
where $\gamma=\beta N+2-\beta$ and
$\beta\in\{1,2,4\}$ is the symmetry index discussed above.
 $J(\{\lambda_{n}\})$ denotes the Jacobian from the matrix to the
eigenvalue space:
\begin{equation}
J(\{\lambda_{n}\})=\prod_{i<j}|\lambda_{j}-\lambda_{i}|^{\beta}.
\label{jacobian}
\end{equation}

We have seen that the RMT is completely identified by its $G/H$
symmetric space. All the irreducible
symmetric spaces of classical type can be classified
with essentially the same techniques used for the Lie algebras.
They fall into 11 classes labelled by the type of root system and by the
multiplicities of the various roots~\cite{helg}.
The DMPK operator $D$ is completely fixed once the information on the
type of root system and the root multiplicities are given. The Dyson
ensambles, which first appeared in the RMT of nuclear energy levels,
(and are also relevant in the RMT of quantum transport based on the
scattering matrix~\cite{fp}) are related to
 the symmetric spaces labelled by the Dynkin diagrams of
type $A_N$. There are exactly three classes of $A_N$ spaces, which are
distinguished by the index
$\beta$ which labels the multiplicity of the roots.
$\beta$ can take the values $1,2$ or $4$ and  exactly coincides with the
symmetry index of the corresponding Random Matrix ensamble.
Instead, the ensambles or random
transfer matrices in which we are interested coincide with
the  symmetric spaces labelled by  Dynkin
diagrams of type $C_N$ (see Tab.1). {\sl It is exactly the
difference between $A_N$ and $C_N$ type of symmetric spaces the
origin of the differences between  RMT based on transfer matrices and
the Wigner-Dyson ones.}

\begin{table}[htb]
\small
\begin{center}
\eqm
\begin{tabular}{|c|c|c|c|c|}
 $G$ & $H$  &  $d$ & $\beta$ & $\eta$ \\
\hline
$Sp(2N,\mbox{\bf R})$ & $U(N)$  & $N(N+1)$ & 1 & 1  \\
$SU(N,N)$ & $SU(N)\otimes SU(N)\otimes U(1)$  & $2N^2$ & 2 & 1  \\
$SO^*(4N)$ & $U(2N)$  & $N(N-1)$ & 4 & 1  \\
$USp(2N,2N)$ & $USp(2N)\otimes USp(2N)$  & $4N^2$ & 4 & 3  \\
$Sp(2N,\mbox{\bf C})$ & $USp(2N)$  & $N(2N+1)$ & 2 & 2  \\
\end{tabular}
\enm
\end{center}
\caption
        {\label{cn}
        Irreducible symmetric spaces of type $C_N$.
        In the first two columns the group $G$ and the maximal
        subgroup $H$ which define the symmetric space. In the third
        column the dimensions $d$ of the space. All these spaces are
        labelled by the Dynkin diagram $C_{N}$. In the fourth column the
        multiplicity $\beta$ of the ordinary roots of $C_N$. In the last
        column the multiplicity $\eta$ of the long root.}
\end{table}

The most relevant feature of the $C_N$ symmetric spaces with respect of
the $A_N$ ones is that they are described by two critical indices
instead of one. This is due to the fact that the $C_N$ Dynkin diagrams
have two types of roots, each with its particular multiplicity. We shall
call this second index $\eta$ in the following. This peculiar feature of
these ensambles was up to now ignored  because in the weakly
localized regime it is only the value of $\beta$ which matters, and also
because the three cases which have been
 studied up to now (the first three lines of tab.1) have the same
value of the second index $\eta=1$, thus leading to the same DMPK
equation (with no explicit $\eta$ dependence).
 However we shall show now that in the insulating regime,
if $both$ the spin-rotation and the time reversal
symmetry are broken, also the fourth ensamble of those listed
in tab.1  could become relevant, and in this case the value of $\eta$
changes.

As a preliminary step to prove this assertion,
let us remind the group properties of the symplectic ensamble of transfer
matrices  (spin-rotation broken, time reversal
conserved). In this case each one of the $I,I',O,O'$ vectors is a
collection of $N$ spinors (one for each channel); each spinor contains
the two spin degrees of freedom. Hence $M$ is in this case a
$4N\times 4N$ complex matrix.

 Flux conservation implies
\eq
|I|^2-|O|^2=|O'|^2-|I'|^2~~~,
\en
which can be written as:
\eq
M^\dagger\Sigma_z M=\Sigma_z
\label{msms}
\en
where
\eq
\Sigma_z=
\left(\begin{array}{cc}\mbox{\bf 1}& \mbox{\bf 0} \\ \mbox{\bf 0}&
\mbox{\bf 1}
 \end{array}\right)\nonumber
\en
where {\bf 1} and {\bf 0} denote the $2N\times2 N$  unit and
zero matrix. Eq.(\ref{msms}) implies that the matrix $M$ belongs to the
unitary group $U(2N,2N)$ group (since it preserves the metric
$\Sigma_z$). Time reversal must then be imposed in two steps. At the
level of the single spinor we must require invariance with respect to
the time reversal operator:
$\theta=i\sigma_y~C$ where $C$ is the complex conjugation operator and
$\sigma_y$ is the Pauli matrix:
\eq
\theta=
\left(\begin{array}{cc} 0 & 1 \\ -1 & 0
 \end{array}\right)~C
\label{m3}
\en
In addition, at the level of the whole matrix $M$ one must impose the
following constraint
\eq
M^*=\Sigma_x M \Sigma_x
\label{m4}
\en
with
\eq
\Sigma_x=
\left(\begin{array}{cc}\mbox{\bf 0}& \mbox{\bf 1} \\ \mbox{\bf 1}&
\mbox{\bf 0}
 \end{array}\right) \nonumber
\en

which, together with eq.(\ref{msms}) implies that $M\in SO^*(4N)$.
Let us now switch on the magnetic field. We have already seen that if
both the constraints (\ref{m3}) and  (\ref{m4}) are eliminated we have
$M\in U(2N,2N)$, namely that $M$ belongs in this case to the unitary
ensamble $\beta=2$ listed in tab.1, {\it but with $N\to 2N$}.
This gives the
standard RMT prediction: $\beta$
changes from 4 to 2 in relation to the level statistics,
as a consequence of the application of a magnetic
field. However for all those observables which are not only related
to the level statistics, but depend more generally on $\gamma$,
 the two changes $\beta: 4\to 2$, $N\to 2N$ compensate and
the model effectively behaves as if $\beta=4$ as in the $\sigma$
model result. This happens in particular for the
localization length $\xi$ in the insulating regime in which we are
interested. In fact we shall show below that $\xi$ is proportional to
$\gamma$. As mentioned above this seems to exclude a negative MC in the
RMT approach, in clear contradiction to the experimental results. To
open the possibility to find a negative MC with RMT, we must make a
further hypothesis.
Since the two constraint (\ref{m3}) and (\ref{m4}) are not on
the same footing, the first being a local and the second a global
property of the electrons, they need not to be simultaneously fulfilled.
In particular
 it could exist a range of parameters (say, for low
magnetic fields) in which only
the second one is eliminated but the first one survives. Physically
this would mean
that locally the Kramers degeneracy is conserved, but globally time-
reversal is broken, with all the known consequences on the
back-scattering paths.
Let us see which are the consequences of this hypothesis.
It is easy to
see that eq.(\ref{m3}) alone implies the conservation of the
antisymmetric metric and that in this case the joint application of
(\ref{msms}) and (\ref{m3}) leads to $M\in USp(2N,2N)$.
This choice of constraints defines the new universality class
anticipated in the title.
Since \eqm $USp(2N,2N)=U(N,N,\mbox{\bf Q})$ \enm
we see that in this case $M$ has a
very natural description in terms of quaternions, being the most general
unitary quaternion matrix which preserves the metric $\Sigma_z$.
Any matrix $M$ of this type can be parametrized as:
\eq
M=\left(\begin{array}{cc} u^{(1)} & \mbox{\bf 0} \\
\mbox{\bf 0} & u^{(2)}  \end{array}\right)
\left(\begin{array}{cc} \sqrt{1+{\bf \Lambda}} &
\sqrt{{\bf \Lambda}} \\
 \sqrt{{\bf \Lambda}} &  \sqrt{1+{\bf \Lambda}}
\end{array}\right)
\left(\begin{array}{cc} u^{(3)} & \mbox{\bf 0} \\
\mbox{\bf 0} & u^{(4)}  \end{array}\right) \nonumber
\en
where the $u^{(i)}$, $(i=1,2,3,4)$ are $N\times N$ quaternion unitary
matrices and
${\bf \Lambda}$  is a  $N\times N$ real, diagonal,
quaternion matrix with non negative elements $\lambda_1,\lambda_2,
\cdots \lambda_N$.
This parametrization is nothing else than the explicit construction of
the symmetric space \eqm $\frac{U(N,N,\mbox{\bf Q})}{U(N,\mbox{\bf Q})\otimes
U(N,\mbox{\bf Q})}$ \enm, which
(through the identification of the quaternion unitary groups as
unitary symplectic groups on the complex) is exactly equivalent to
the one listed in the fourth
line of tab.1 .
Let us now generalize
the DMPK equation to take into account this new critical index.
The (generalized) DMPK equation can be
obtained as a particular case of the general Calogero-Sutherland
construction (see~\cite{c1,op}). The result is:
\begin{equation}
\frac{\partial P}{\partial s}~=
\frac{2}{\tilde\gamma}\sum_{i=1}^{N}
\frac{\partial}{\partial\lambda_{i}}\lambda_{i}(1+\lambda_{i})
\tilde J\frac{\partial}{\partial\lambda_{i}}\tilde{J}^{-1}~P
\label{DMPK2}
\end{equation}
where
\begin{equation}
\tilde{J}(\{\lambda_{n}\})=\prod_{i<j}|\lambda_{j}-\lambda_{i}|^{\beta}
\prod_i [\lambda_{i}(1+\lambda_{i})]^{\frac{\eta-1}{2}}~~,
\label{j2}
\end{equation}
$\tilde\gamma=\frac{2}{\eta+1}(\beta (N-1) +1+\eta)$
and the indices $\beta$ and $\eta$ are listed in tab.1.

The DMPK equation greatly simplifies if one chooses a new set of
variables $\{x_n\}$ related to the eigenvalues $\{\lambda_i\}$ by
$\lambda_n=\sinh^2x_n$. Then the solution can be written in terms of so
called ``zonal spherical functions'' $\Phi_k(x)$, $x=\{x_1,\cdots,x_N\}$,
$k=\{k_1,\cdots,k_N\}$ (which depend on the chosen symmetric space).

\begin{equation}
P(\{x_n\},s)=[\xi(x)]^2 \int dk
 e^{-\frac{k^2}{2\tilde\gamma}s}
\frac{ \Phi_k(x)}{|\Delta(k)|^2}F(k)
\label{ss3}
\end{equation}

where
\begin{equation}
\xi(x)=\prod_{i<j}|\sinh^{2}x_{j}-\sinh^{2}x_{i}|^{
\frac{\beta}{2}}
\prod_{i}|\sinh 2x_{i}|^{\frac{\eta}{2}}~~~;
\label{BR2}
\end{equation}

\eq
F(k)= \prod_{j}
k_j\tanh(\frac{\pi k_j}{2})(k_j^2+1)^{\frac{\eta-1}{2}}
\en
\begin{equation}
|\Delta(k)|^2=\prod_{m<j}\left \vert\frac{\Gamma\left(
i\frac{k_m-k_j}{2}\right)\Gamma\left(
i\frac{k_m+k_j}{2}\right)}{\Gamma\left(\frac{\beta}{2}+
i\frac{k_m-k_j}{2}\right)\Gamma\left(\frac{\beta}{2}+
i\frac{k_m+k_j}{2}\right)}\right\vert^2
\label{ss2b}
\end{equation}
and $\Gamma$ denotes the Euler gamma function.

In the insulating regime $(k\ll 1)$ we can use the asymptotic expansion
described in~\cite{c1} for the zonal spherical function. In this way
the  integration over $k$ can be done explicitly and gives~\cite{c1}:

\begin{eqnarray}
P(\{x_{n}\},s)&=&\prod_{i<j}\left\vert\sinh^{2}x_{j}-\sinh^{2
}x_{i}\right\vert^{\frac{\beta}{2}}\left[
(x_{j}^{2}-x_{i}^{2})\right]\nonumber\\
&\times &\prod_{i}\left[\exp(-x_{i}^{2}\tilde\gamma/(2s))x_{i}(\sinh
2x_{i})^{\eta/2}\right].
\label{f1}
\end{eqnarray}

 Ordering the
$x_{n}$'s from small to large and using the fact that in this regime
 $1\ll x_{1}\ll x_{2}\ll\cdots\ll x_{N}$ we can
approximate the eigenvalue distribution as follows:
\begin{equation}
P(\{x_{n}\},s)=\prod_{i=1}^{N}
\exp\left[-(\tilde\gamma/(2s))(x_{i}-\bar{x}_{i})^{2}\right].
\label{IR2}
\end{equation}
where $\bar{x}_{n}=\frac{s}{\tilde\gamma}(\eta+\beta(n-1))$.
In the strongly localized limit the conductance is dominated by the
first eigenvalue of the transmission matrix:
\eq
<\log{g}>\equiv-\frac{2L}{\xi}\sim\log{T_1}\sim -
2\bar x_1=\frac{2s\eta}{\tilde\gamma}.
\label{g1}
\en
and hence:
\eq
\xi=\frac{2l}{\eta(\eta+1)}[\beta(N-1)+1+\eta]
\en

Setting now the proper values of $\beta$, $N$ and $\eta$ for the various
regimes, and taking the large $N$ limit, we see, as anticipated, that if
both time reversal and Kramers degeneracy are broken, namely if we move
from
$\frac{SO^*(4N)}{U(2N)}\to
\frac{SU(2N,2N)}{SU(2N)\otimes SU(2N)\otimes U(1)}$
then $\xi$ does not change, hence $\xi(H)/\xi=1$ and the MC is 0.
If we assume that Kramers degeneracy is conserved then we move from
$\frac{SO^*(4N)}{U(2N)}\to
\frac{USp(2N,2N)}{USp(2N)\otimes USp(2N)}$ and  $\xi(H)/\xi=1/6$ thus
implying a negative MC.

Needless to say that the negative MC measured in
the experiments could also simply indicate a failure of the RMT approach
outside the metallic phase. Moreover one must be aware of the fact
that all the above results are constrained to the quasi
one-dimensional limit which underlies the DMPK equation.
Nevertheless, it
would be very interesting to test the possible existence of this
new universality class. In this respect
a stringent and simple test is given by the ratio between the mean
and the variance
of $\log g$. This ratio is easily extracted from
eq.s(\ref{IR2},\ref{g1}) and turns out to be
\eq
\frac{var~(\log{g})}{<\log{g}>}=\frac{2}{\eta}.
\en
This observable gives a direct measure of $\eta$ and could play in the
insulating regime the role which is played in the metallic one by the
universal conductance fluctuations for $\beta$.
We hope that our results will
stimulate new experiments in order clarify this issue.

\end{document}